\documentclass[osajnl,twocolumn,showpacs,superscriptaddress,10pt]{revtex4-1}
\usepackage[colorlinks,citecolor=blue,linkcolor=red,dvipdfm]{hyperref}
\usepackage{amsmath,amssymb,graphicx}

\begin{document}

\title{Two-dimensional intra-band solitons in lattice potentials with local
defects and self-focusing nonlinearity}
\author{Jianhua Zeng}
\email{Email: zengjianhua1981@gmail.com}
\affiliation{State Key Laboratory of Low Dimensional Quantum Physics,
Department of Physics, Tsinghua University, Beijing 100084, China}
\author{Boris A. Malomed}
\email{Email: malomed@post.tau.ac.il} 
\affiliation{Department of Physical Electronics, School of
Electrical Engineering, Faculty of Engineering, Tel Aviv University,
Tel Aviv 69978, Israel}

\begin{abstract}
It is commonly known that stable bright solitons in periodic
potentials, which represent gratings in photonics/plasmonics, or
optical lattices in quantum gases, exist either in the spectral
semi-infinite gap (SIG) or in finite bandgaps. Using numerical
methods, we demonstrate that, under the action of the cubic
self-focusing nonlinearity, defects in the form of ``holes" in
two-dimensional (2D) lattices support continuous families of 2D
solitons \textit{embedded} into the first two Bloch bands of the
respective linear spectrum, where solitons normally do not exist.
The two families of the \textit{embedded defect solitons} (EDSs) are
found to be continuously linked by the branch of \textit{gap defect
solitons} (GDSs) populating the first finite bandgap. Further, the
EDS branch traversing the first band links the GDS family with the
branch of regular defect-supported solitons populating the SIG.
Thus, we construct a continuous chain of regular, embedded, and
gap-mode solitons (``superfamily") threading the entire bandgap
structure considered here. The EDSs are stable in the first Bloch
band, and partly stable in the second. They exist with the norm
exceeding a minimum value, hence they do not originate from linear
defect modes. Further, we demonstrate that double, triple and
quadruple lattice defects support stable dipole-mode solitons and
vortices.

\end{abstract}

\maketitle

\section{Introduction}

A powerful toolbox for controlling the dynamics of photonic, plasmonic, and
atomic waves is based on the use of periodic potentials, which can be
induced by photonic lattices and crystals in optics \cite{PC}, by gratings
built into plasmonic waveguides (see original works \cite%
{plasmon1,plasmon2,plasmon3,plasmon4,plasmon5,plasmon6,
plasmon7,plasmon8,plasmon9,plasmon10,plasmon11,plasmon12} and review \cite%
{plasmon-review}), and by optical lattices (OLs) in Bose-Einstein
condensates (BECs) \cite{BEC,reviews-1Da,reviews-1Db} or degenerate Fermi
gases \cite{Fermi}. Acting in the combination with material nonlinearity,
periodic potentials give rise to various species of solitons. These include
ordinary ones, existing in the semi-infinite gap (SIG) of the corresponding
linear spectrum, and gap solitons populating finite bandgaps, see reviews
\cite{reviews-1Da,reviews-1Db,review-multiD,Barcelona-review} and book \cite%
{Peli}. Extended localized states, in the form of truncated nonlinear Bloch
waves, which appear in finite bandgaps, have been observed and modeled too
\cite{G-waves1,G-waves2,G-waves3}.

On the other hand, a specific species of solitary waves is known in
the form of \textit{embedded solitons}, which exist,
counter-intuitively, inside spectral Bloch bands populated by
radiation modes \cite{ES1}-\cite{Yang}. Although embedded solitons
cannot exist, generically, in continuous families, due to the
resonance with radiation waves, isolated embedded modes are
possible, for which the rate of the decay into radiation vanishes.
Under more specific conditions, continuous families of embedded
solitons were constructed too \cite{ES10,ES11,ES12,ES13}. Very
recently, the existence of solitons of this type was demonstrated in
a model combining the quadratic ($\chi ^{(2)}$) nonlinearity and a
complex lattice potential, whose imaginary part is subject to the
condition of the $\mathcal{PT}$ symmetry, representing
symmetrically placed and mutually balanced local gain and loss \cite%
{Frederico}.While embedded solitons have been studied in some detail in
basic one-dimensional (1D) models, it is still an open question to find them
in Bloch bands of 2D periodic potentials (stable intra-band solitons were
reported in a 2D potential which combines a periodic lattice in one
direction and a harmonic-oscillator trap acting perpendicular to the lattice
\cite{Yang}).

In addition to the studies of ideal lattices, it has been demonstrated that
lattice defects may also be used in optical devices, such as integrated
circuits \cite{PC-defect1a,PC-defect1b,PC-defect2}, microcavities \cite%
{micro}, and defect-mode lasers \cite{laser}. Point and line defects help
trapping and guiding light flows in structured photonic media \cite%
{PC-defect31}-\cite{PC-defect4b}. In particular, the bandgap
guidance of light by defects in 1D and 2D photonic lattices has been
realized in Refs. \cite{PL1,PL2,PL3}. Defect-affected transmission
of light was also considered in 2D photonic quasicrystals \cite%
{quasicrystals1,quasicrystals2}. The dynamics of solitons under the
action of defects in discrete nonlinear-Schr\"{o}dinger lattices has
been studied too \cite{DD1}-\cite{DD6}. The dynamics of BEC trapped
in OL potentials may also be strongly affected by defects
\cite{BAM1}-\cite{defect-TB}.

The above-mentioned settings are based on the cubic nonlinearity.
Solitons supported by local defects in the form of the localized
$\chi ^{(2)}$ nonlinearity have been studied in Refs.
\cite{chi2a,chi2b,chi2c}. Very recently, solitons pinned to local
$\mathcal{PT}$-symmetric defects, inserted into the 1D medium with
the uniform $\chi ^{(2)}$ nonlinearity, were considered too
\cite{Konotop}. On the other hand, a specific manifestation of the
transmission of light \cite{Anderson1}-\cite{Anderson8} and matter
waves \cite{Anderson-BEC1}-\cite{Anderson-BEC4} controlled by random
sets of defects is the Anderson localization in the transverse
plane, which does not require nonlinearity.

More recently, a great deal of attention has been drawn by soliton dynamics
in nonlinear lattices, i.e., structures periodically modulating the local
nonlinearity \cite{NL_review,NL_Oursa,NL_Oursb}. In this context, defects
can be naturally considered as well. In particular, experimental techniques
have been developed for the local control of local nonlinearity in BEC (via
the Feshbach resonance \cite{MagnLatt,MagnLatt2}), and in photonic crystals
\cite{NL_PC,NL_PC2}, which allows one to design desirable nonlinear defects.

In this work we report numerical analysis of 2D solitons in lattice
potentials with local defects, under the action of the self-focusing
nonlinearity. The model is formulated in Section II. The search for such
soliton modes is suggested by previously published results demonstrating the
existence of linear defect modes in photonic crystals \cite{PC-defect1a}-%
\cite{PC-defect4b}, including 2D modes \cite{2D-mode}, as well as in
ultrasonic crystals \cite{sonic} and solid-state lattices (see Refs. \cite%
{solid-state,solid-state2,solid-state3,solid-state4} and references
therein), although the solitons that we present below are not actually
related to linear defect modes. In Section III, we report numerical results
which demonstrate the existence of a new type of embedded solitons, alias
\textit{intra-band} solitons, \textit{viz}., \emph{continuous} families of
2D localized modes (unlike isolated solutions found previously in typical
embedded-soliton models), which are embedded into the first and second Bloch
bands of the linear spectrum, in terms of the propagation constant, and
pinned to the defect, as concerns the spatial location of the modes. We name
them embedded defect solitons (EDSs). It is relevant to stress that, for
modes pinned to local lattice defects, it makes sense to identify the
location of their propagation constant with respect to the bandgap spectrum
of the infinite uniform OL, as the latter determines the spectrum of
radiation waves into which the localized mode may decay.

As mentioned above, the EDSs \emph{do not} emerge from a
continuation of linear localized defect modes, as they exist with
the norm exceeding a finite minimum value (and the propagation
constant of linear defect modes cannot be located inside a Bloch
band). By means of systematic direct simulations, the EDSs are found
to be fully stable in the first band and in a part of the second.
Together with the regular defect solitons populating the SIG, and
gap defect solitons (GDSs) nested in the first finite bandgap (in
the spectral slot between the first and second Bloch bands), which
are pinned to the same defect, the intra-band EDSs form a continuous
``superfamily" threading the entire band structure. The stability of
the GDSs supported by the \emph{self-attractive} nonlinearity is a
remarkable finding, as the previous analysis of gap solitons in
defect-free 2D lattices was usually restricted to the case of the
self-repulsion, assuming that they would be unstable in the opposite
case \cite{review-multiD, repulsive-2D,repulsive-2D2} (gap solitons
bifurcating under the action of the self-attractive nonlinearity
from edges of the adjacent Bloch bands into the first finite bandgap
were constructed in Ref. \cite{Yang2D}, but their stability was not
studied there; in fact, it was recently found that, while 2D
fundamental gap solitons are completely unstable in the first finite
bandgap under the self-attractive nonlinearity, a family of
\textit{dipole-mode} gap solitons are stable in a part of the first
bandgap in the same case, provided that the depth of the lattice
potential exceeds a certain threshold value \cite{Nir}).

Further, in Section IV we demonstrate the existence of stable EDS and GDS
modes with the dipole structure, in the lattice with double and triple local
defects, and of vortices supported by quadruple defects. The paper is
concluded by Section V.

\section{The model}

\begin{figure}[tbp]
\begin{center}
\includegraphics[height=6.cm]{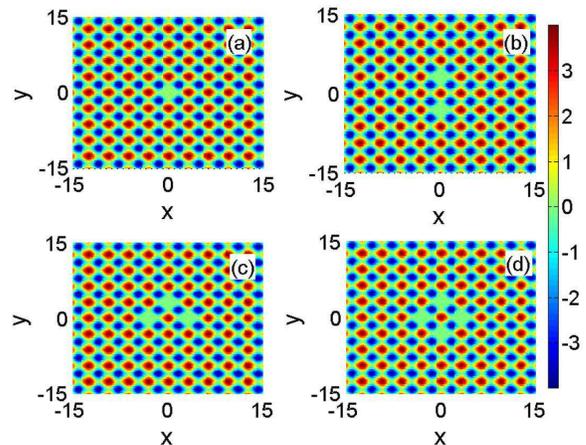}
\end{center}
\caption{(Color online) Contour plots of the lattice potential with the
single (a), double (b), triple (c), and quadruple (d) defects, at $\protect%
\varepsilon =2$.}
\label{Fig1}
\end{figure}

The analysis is based on the 2D nonlinear Schr\"{o}dinger/Gross-Pitaevskii
equation for the mean-field BEC wave function, or the amplitude of the
electromagnetic wave propagating in a photonic lattice, $\psi (x,y,z)$,
\begin{equation}
i\psi _{z}=-(1/2)\nabla ^{2}\psi -[1-\tilde{\delta}(x,y)]V_{\mathrm{OL}}\psi
-|\psi |^{2}\psi  \label{GPE}
\end{equation}%
(in BEC, propagation distance $z$ is replaced by time $t$), with the
self-attractive nonlinearity and $\nabla ^{2}=\partial _{x}^{2}+\partial
_{y}^{2}$. In the absence of the defect, the lattice potential with depth $%
2\varepsilon $ and period normalized to be $\pi $ is taken in the
usual form, $V_{\mathrm{OL}}=\varepsilon \lbrack \cos (2x)+\cos
(2y)]$. The defect is a ``hole" of radius $r_{0}=1.2$ in the
potential, which
is accounted for by $\tilde{\delta}(x,y)=1$ at $x^{2}+y^{2}<r_{0}^{2}$, and $%
\tilde{\delta}(x,y)\equiv 0$ at $x^{2}+y^{2}\geq r_{0}^{2}$ in Eq. (\ref{GPE}%
). Generic results can be adequately demonstrated for this shape$\
$of the defect with $\varepsilon =2$, which is fixed below (in fact,
fully stable soliton families could be found in the interval of
$1.1\leq r_{0}\leq 1.3$, as well as for different shapes of the
``hole"). The corresponding contour plots of the OL potential with
the solitary or compound defects are displayed in Fig. \ref{Fig1}.

Stationary solutions to Eq. (\ref{GPE}) with propagation constant $-\mu $
(or chemical potential $\mu $, in terms of BEC) are sought for as $\psi
=\phi (x,y)\exp (-i\mu z)$, with function $\phi (x,y)$ obeying the
stationary equation,
\begin{equation}
\mu \phi =-(1/2)\nabla ^{2}\phi -[1-\delta (x,y)]V_{\mathrm{OL}}\left(
x,y\right) \phi -|\phi |^{2}\phi .  \label{phi}
\end{equation}%
While fundamental and dipole-mode solitons are described by real solutions
for $\phi \left( x,y\right) $, vortices (that can be supported by quadruple
defects, see below) naturally correspond to complex ones.

Figure \ref{Fig2} displays the linear spectrum of Eq. (\ref{GPE}), which has
been produced by a numerical solution of the linearized version of Eq. (\ref%
{phi}). It is seen that, for given values of the parameters, a growing
number of defect-induced isolated eigenvalues appear in the second finite
bandgap with the increase in the size of the defect (single $\rightarrow $
double $\rightarrow $ triple), while the first bandgap remains unaffected.
Below, we focus not on quite obvious quasi-linear defect modes corresponding
to these isolated eigenvalues, but rather on localized states which have
\emph{no linear limit}, but may be created by the defects in the first
finite bandgap, as well as in the two Bloch bands adjacent to it. These
states were constructed as numerical solutions of Eq. (\ref{phi}) by means
of the Newton's method in the domain of size $30\times 30$, covered by a
grid of $192\times 192$ points. The initial guess was the isotropic
Gaussian, $\phi \left( x,y\right) =A\exp \left[ -\left( x^{2}+y^{2}\right)
/\left( 2W^{2}\right) \right] $, with amplitude $A$, width $W$, and total
power (or number of atoms, in the BEC), $N\equiv \int \int \phi ^{2}\left(
x,y\right) dxdy=\pi \left( AW\right) ^{2}$. The stability of the solutions
was subsequently tested in direct simulations of perturbed evolution in the
framework of Eq. (\ref{GPE}), using the same numerical domain with absorbing
boundary conditions.
\begin{figure}[tbp]
\begin{center}
\includegraphics[height=6.cm]{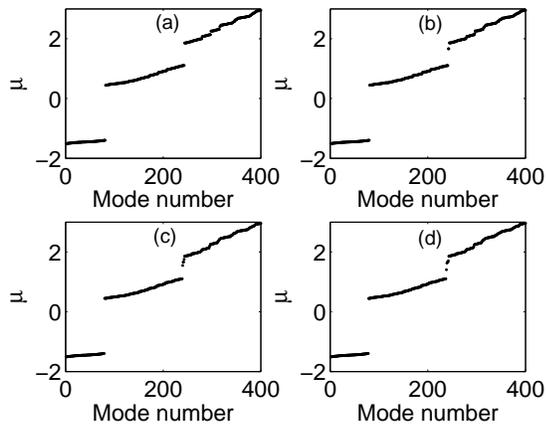}
\end{center}
\caption{ Linear spectra of the perfect OL (a), and OLs with the single (b),
double (c), and triple (d) defects displayed in Fig. \protect\ref{Fig1}.
Isolated points emerging in the second finite bandgap represent the
corresponding linear defect modes.}
\label{Fig2}
\end{figure}

\section{Results of the analysis for the single defect}

\begin{figure}[tbp]
\begin{center}
\includegraphics[height=4.cm]{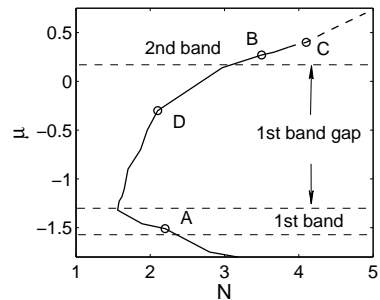}
\end{center}
\caption{ The propagation constant ($-\protect\mu $) versus the total power
norm, for the families of solitons pinned to the single defect [it is shown
in Fig. \protect\ref{Fig1}(a)], which were found inside the first and second
Bloch bands, as well as in the first finite bandgap between them. The solid
and dashed curves indicate stable and unstable solutions, respectively.
Typical examples of the pinned solitons, corresponding to the marked points,
are shown below in Figs. \protect\ref{Fig4} and \protect\ref{Fig5}. Some
irregularity of the curves is explained by the fact that they were plotted
with interval $\Delta \protect\mu =0.1$.}
\label{Fig3}
\end{figure}

As shown in Fig. \ref{Fig3}, the numerical solution reveals the existence of
continuous families of localized modes (fundamental EDSs), pinned to the
single defect [the one shown in Fig. \ref{Fig1}(a)], inside the first and
second Bloch bands. The EDS families are continuously linked to the family
of GDSs, attached to the same defect, whose propagation constant belongs to
the first finite bandgap. The EDS family in the first Bloch band starts
exactly at its boundary with the SIG, where it is linked to the family of
stable regular solitons pinned to the defect. The solitons of the latter
type are not presented here in detail, as their existence and properties are
quite obvious.

It is observed that the combined family of the pinned solitons in Fig. \ref%
{Fig3} exists above a finite minimum value of the total power, $N_{\min
}=1.56$, which exactly corresponds to the boundary between the first Bloch
band and the first finite bandgap. This means, as said above, that the
family of the localized modes supported by the defect does not emerge as a
nonlinear continuation of any linear defect mode, as the latter would
correspond to $N\rightarrow 0$.

Direct simulations demonstrates that the EDS family is completely stable in
the first Bloch band, while the EDS becomes unstable in the second band
beyond a certain critical point. It is worthy to note that the stability of
the intra-band solitons supported by the defect in the first Bloch band
agrees with the Vakhitov-Kolokolov (VK) criterion, which is relevant to
localized modes in the case of the self-focusing nonlinearity \cite{refVK}, $%
\partial \mu /\partial {N}<0$, while the EDS branch features $\partial \mu
/\partial {N}>0$ in the second band, where the modes are stable in a limited
region (strictly speaking, this may be a region of a very weak instability,
which is, nevertheless, tantamount to stability in terms of a possible
experiment). Of course, there is no direct proof of the applicability of the
VK criterion to the present model.

\begin{figure}[tbp]
\begin{center}
\includegraphics[height=9.cm]{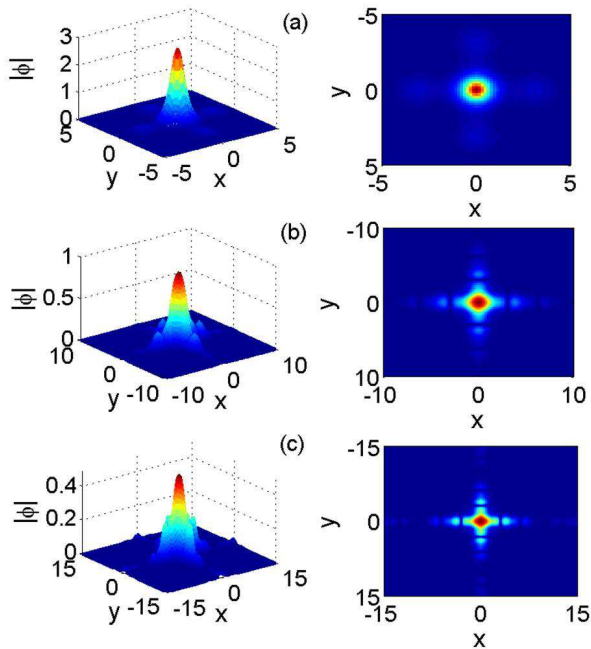}
\end{center}
\caption{(Color online) Examples of the intra-band (\textit{embedded})
defect solitons, EDSs, found in the first (a) and second (b,c) Bloch bands,
which correspond to points A and B,C, respectively, in Fig. \protect\ref%
{Fig3}. The shapes of the modes are shown by means of 3D images (left) and
contour plots (right). The mode in panels (a) and (b) are stable, while the
one in (c) is unstable.}
\label{Fig4}
\end{figure}

Typical examples of stable EDSs in the first and second Bloch bands, found
at points marked (A) and (B) in Fig. \ref{Fig3}, along with an example of an
unstable EDS corresponding to point C, are displayed in Fig. \ref{Fig4}.
Throughout the first band, the shape of the solitons is very similar to that
displayed in Fig. \ref{Fig4}(a). On the other hand, the localized modes
feature a more complex structure, with pronounced side peaks, in the second
band [see Figs. \ref{Fig4}(b,c)], in accordance with the fact that the shape
of Bloch modes is more complex too in the same band. In fact, the extended
tail of the soliton shown in Fig. \ref{Fig4}(c) initiates the onset of the
instability of this soliton. In direct simulations, unstable modes decay
into radiation (not shown here in detail).

To the best of our knowledge, the current model presents the first example
of nonlinear localized modes found inside Bloch bands of the spectrum
induced by the 2D lattice (in Ref. \cite{Yang}, which was dealing with 2D
solitons of the embedded type, the lattice was one-dimensional). In fact,
the EDS family threading the first Bloch band plays the role of a missing
link between the two well-known types of solitons existing in the lattice
potential, \textit{viz}., regular ones populating the SIG, and gap solitons
in the first finite bandgap (in the present case, they all are pinned to the
single defect).

As concerns the family of GDSs found inside the first finite bandgap, which
links the intra-band EDS families populating the two adjacent Bloch bands
(see Fig. \ref{Fig3}), our numerical tests have demonstrated that these
solitons, being pinned to the defect, are stable only in the case of the
self-focusing nonlinearity [as set in Eq. (\ref{GPE})], on the contrary to
gap solitons in perfect lattices, which are usually assumed to be stable
solely in the case of the self-defocusing \cite{review-multiD, repulsive-2D}%
. The case of the self-defocusing nonlinearity was explored too, and
it was concluded that the pinned GDSs are unstable in that case (not
displayed here in detail). The latter finding may be explained by
the fact that the effective mass of the gap soliton supported by the
self-defocusing nonlinearity is \emph{negative}, thus reversing the
character of the soliton-defect interaction, and making unstable the
bound state of the gap soliton pinned to the attractive defect. In
line with this argument, additional analysis demonstrates that the
local defect of the opposite sign, in comparison with the one
considered here, gives rise to stable pinned states of gap solitons
under the self-defocusing nonlinearity, and unstable states in the
case of the self-focusing. The negative mass of gap solitons
explains a number of other counter-intuitive dynamical effects
featured by these modes \cite{negative}-\cite{negative4}. Numerical
results also demonstrate that the self-defocusing nonlinearity
\emph{does not} give rise to solitons that would be embedded into
Bloch bands and pinned to the defect.

\begin{figure}[tbp]
\begin{center}
\includegraphics[height=4.cm]{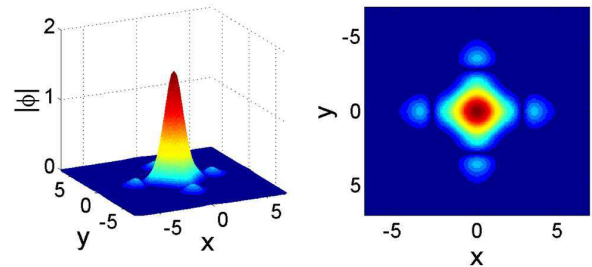}
\end{center}
\caption{(Color online) An example of a stable gap soliton pinned to the
single defect, found in the first bandgap (at point D marked in Fig. \protect
\ref{Fig3}) under the self-focusing nonlinearity.}
\label{Fig5}
\end{figure}

A typical example of the stable pinned GDS, found in the first finite
bandgap in the case of the self-focusing nonlinearity, which corresponds to
point D in Fig. \ref{Fig3}, is presented in Fig. \ref{Fig5}. Finally, our
numerical simulations suggest that stable GDSs pinned to the defect do not
exist in the relatively narrow second finite bandgap (see Fig. \ref{Fig2} ),
in agreement with the general trend of gap solitons to be unstable in the
second bandgap \cite{reviews-1Da,reviews-1Db,Peli}.

\section{Results for multiple defects: stable dipoles and vortices}

\begin{figure}[tbp]
\begin{center}
\includegraphics[height=4.cm]{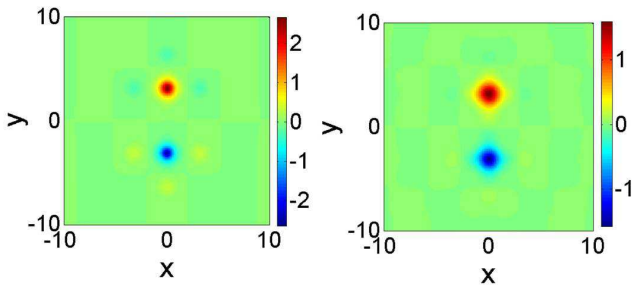}
\end{center}
\caption{(Color online) Examples of stable dipole-mode solitons supported by
the double defect, shown in Fig. \protect\ref{Fig1}(b), in the first Bloch
band (left, with $\protect\mu =-1.4,N=3.9$), and in the first finite bandgap
(right, with $\protect\mu =-0.3,N=5.4$).}
\label{Fig6}
\end{figure}

We have extended the above analysis for the double, triple, and quadruple
lattice defects, which are shown in Fig. \ref{Fig1}(b,c,d). While the
solitary defect may only support fundamental solitons of both the EDS and
GDS types, as shown above, the double defects readily give rise to stable
dipole-mode bound states of two solitons with opposite signs. Examples of
such modes, found in the first Bloch band and in the first finite bandgap
(i.e., of the EDS and GDS types, respectively), are shown in Fig. \ref{Fig6}%
.
\begin{figure}[tbp]
\begin{center}
\includegraphics[height=4.cm]{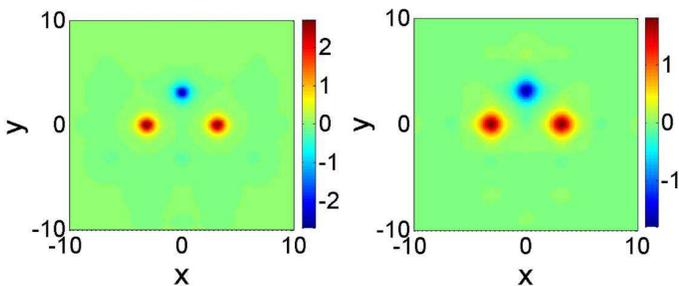}
\end{center}
\caption{(Color online) Examples of stable dipole-mode solitons supported by
the triple triangular defect, shown in Fig. \protect\ref{Fig1}(c), in the
first Bloch band (left, $\protect\mu =-1.45,N=6.3$), and in the first finite
bandgap (right, $\protect\mu =-0.5,N=6.9$).}
\label{Fig7}
\end{figure}

The triangular defects [see Fig. \ref{Fig1}(c)] also support stable
triangularly shaped dipole modes, see examples displayed in Fig. \ref{Fig7}.
Further, the quadruple defects [see Fig. \ref{Fig1}(d)] create stable
solitary vortices, characteristic examples of which are shown in Fig. \ref%
{Fig8}. As is typical for vortices supported by lattice potentials \cite%
{vortex,vortex2,NL_Oursb, review-multiD}, they are built as rhombic sets of
four intensity peaks, with a nearly empty site in the center (therefore this
type of vortices is often called onsite-centered), and phase shifts $\pi /2$
between the peaks, which corresponds to the global phase circulation of $%
2\pi $ (i.e., topological charge $1$). As indicated in Fig. \ref{Fig9},
direct simulations demonstrate that the pinned vortices are unstable in the
second Bloch band, being stable elsewhere.
\begin{figure}[tbp]
\begin{center}
\includegraphics[height=6.cm]{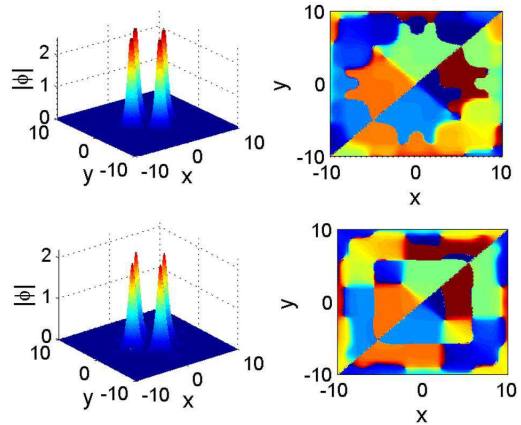}
\end{center}
\caption{(Color online) Examples of stable vortices with topological charge $%
1$, supported by the quadruple defect inside the first Bloch band (top, $%
\protect\mu =-1.5$ and $N=8.8$), and in the first bandgap (bottom, $\protect%
\mu =-0.8$ and $N=7.2$ ). The contour plots show the phase distribution in
the vortices.}
\label{Fig8}
\end{figure}
\begin{figure}[tbp]
\begin{center}
\includegraphics[height=4.cm]{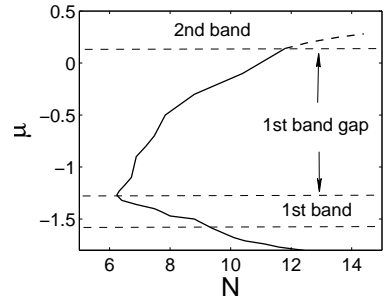}
\end{center}
\caption{Curve $\protect\mu (N)$ for the vortices with topological charge $1$%
, pinned to the quadrupole lattice defect. The solid and dashed curves show
stable and unstable solutions, respectively.}
\label{Fig9}
\end{figure}

\section{Conclusion}

In this work, we have studied localized states which can be supported by
single, double, triple, and quadruple defects of 2D lattice potentials in
the self-focusing medium. The model describes lattices in nonlinear photonic
media and BEC trapped in the OL. The first result is that the single defect
supports families of intra-band (\textit{embedded}) solitons, in the first
and second Bloch bands. These modes exist above the minimum value of the
norm, i.e., they do not emerge from any linear defect mode. Direct
simulations demonstrate stability of the family in the first Bloch band, and
in a part of the second band. The branch of the stable pinned solitons
embedded into the first band links families of regular and gap solitons
hosted by the semi-infinite and first finite bandgaps, respectively, and
pinned to the same defect. The family of the gap solitons, pinned to the
defect, is stable, despite the fact that the nonlinearity is
self-attractive, while it is usually assumed that 2D fundamental gap
solitons may be stable only under the self-repulsion (in uniform lattices).
Finally, it has been found that the double and triple defects support stable
dipole solitons, in the first Bloch band and in the first finite bandgap
alike, and, similarly, the quadruple defect creates stable vortex solitons.

This work can be naturally extended in other directions---in
particular, to the model with a rectilinear defect, as an elongated
version of the double one. Search for stable vortex solitons with
higher topological charges, and, possibly, ``supervortices" (vortex
rings built of compact localized vortices) \cite{super,super2} may
be relevant too. On the other hand, it may be interesting to
consider defects in complex lattices, whose imaginary part is
subject to the condition of the $\mathcal{PT}$ symmetry. 2D gap
solitons in uniform $\mathcal{PT}$-symmetric lattices were found
recently, but they are unstable \cite{PT}. Stable 1D solitons pinned
to defects in $\mathcal{PT}$-symmetric lattices, embedded into the \
medium with the self-defocusing cubic nonlinearity, were reported
very recently in Ref. \cite{recent}. In this connection, it is also
relevant to mention that 1D and 2D gap solitons can be readily made
stable in a more general Ginzburg-Landau system, which does not
impose the balance condition on the gain and dissipation
\cite{GL,GL2}.

\section{Acknowledgments}

J.Z. acknowledges support from the Natural Science Foundation of China
(Project No. 11204151) and China Postdoctoral Science Foundation
(2012M520238).

\end{document}